\documentclass[floats,floatfix,showpacs,amssymb,prd,twocolumn,superscriptaddress,nofootinbib,reprint]{revtex4-2}

\usepackage{graphicx}
\usepackage{dcolumn}
\usepackage{bm}
\usepackage{amsmath}

\usepackage{color}
\usepackage{ulem}

\newcommand{\aap}{Astron. \& Astrophys.}
\def\vev#1{{\left\langle #1 \right \rangle}}

\begin{document}

\preprint{APS/123-QED}

\title[Tensions in CMB quadrupoles]{Hints of tensions in the cosmic microwave background temperature and polarization quadrupoles}

\author{Jahmour J. Givans}
\email{Email: jgivans@flatironinstitute.org}
\affiliation{%
 Center for Computational Astrophysics, Flatiron Institute, 162 5th Ave, New York, NY 10010, USA
}%
\affiliation{%
 Department of Astrophysical Sciences, Peyton Hall,  Princeton University, Princeton, NJ 08544, USA
}%

\author{Marc Kamionkowski}
 
\affiliation{
 William H.\ Miller III Department of Physics \& Astronomy, Johns Hopkins University, 3400 N.\ Charles St., Baltimore, MD 21218, USA
}%

\date{\today}

\begin{abstract}
The large-angular-scale falloff in the autocorrelation function for the cosmic microwave background (CMB) temperature has long intrigued cosmologists and fueled speculation about suppressed superhorizon power.  Here we highlight an inconsistency between the temperature quadrupole and the more recently obtained E-mode polarization quadrupole from Planck PR3.  The temperature quadrupole arises primarily at the CMB surface of last scatter, while the polarization primarily from the epoch of reionization, but the two still probe comparable distance scales.  Although the temperature quadrupole is intriguingly low (much greater than a $1\sigma$ fluctuation) compared with that expected in the standard $\Lambda$CDM cosmological model, the polarization quadrupole turns out to be somewhat high, at the $1\sigma$ level.  We calculate the joint probability distribution function for both and find a slight tension: the observed pair of quadrupoles is inconsistent at a $2.3\sigma$ confidence level. The problem is robust to simple changes to the cosmological model.  If the high polarization quadrupole survives further scrutiny, then this result disfavors, at comparable significance, new superhorizon physics.  The full-sky coverage and pristine foreground subtraction of the LiteBIRD satellite will be ideal to help resolve this question.
\end{abstract}

\maketitle


\section{Introduction}

The dropoff in the cosmic microwave background (CMB) temperature autocorrelation function at the largest angular scales has intrigued cosmologists ever since it was first seen in the Cosmic Background Explorer (COBE) Differential Microwave Radiometer (DMR) maps \citep{Bennett:1996ce}.  This dropoff is seen in the correlation function at angular separations $\Delta \theta \gtrsim 60^\circ$.  In harmonic space, it is manifest as an unusually low temperature quadrupole ($C_2$) \citep{Bennett:1996ce,WMAP:2012fli,Planck:2018vyg}; it is about one fifth the value expected from extrapolation of the power spectrum ($C_l$) at higher multipoles $l$.

This fluctuation is nothing to lose sleep over, especially if you take into account the look-elsewhere effect (of the thousands of multipole moments that have been measured, you'd expect a few to stray from the curve).  But on the other hand, the quadrupole probes the largest observationally accessible distance scales in the Universe.  Maybe there's something happening at larger, superhorizon, distances, and this is just the tip of the iceberg?  This possibility has fueled a number of ideas for new superhorizon physics \citep{Boyanovsky:2006qi,Chen:2008wn,Ramirez:2011kk,Dudas:2012vv,Contaldi:2003zv,Destri:2009hn,Cicoli:2014bja,Kitazawa:2015uda,Bousso:2013uia,Chluba:2015bqa,COMPACT:2022gbl,Braglia:2020fms,Ragavendra:2020old}, and motivates the search for other ways to access information on these superhorizon scales.

One avenue of investigation \citep{WMAP:2003elm,Copi:2008hw,Copi:2010na,Planck:2019evm} has involved study of the statistical significance of the dearth of large-angle correlations, the possibility of cross-correlations between different multipole moments that affect the interpretation.  The conclusion, though, is that cosmic variance, the sample variance that arises from the finite number of $\sim60^\circ$ patches on the sky, prevents the statistical significance to rise much beyond the $\sim2\sigma$ level. Another avenue, though, is to seek other observables that probe the largest distance scales and that may thus provide complementary information.  Ideas along these lines include the additional information provided by lensing maps \citep{Yoho:2013tta} and large-scale polarization fluctuations in the CMB \citep{Copi:2013zja,Yoho:2015bla}.  The latter was considered by the Planck Collaboration in two different ways: Ref.~\cite{Planck:2019evm} generalized the analysis of the temperature autocorrelation function to include temperature-polarization cross-correlation and polarization autocorrelations---the conclusions of prior work were essentially unchanged. Ref.~\cite{2020A&A...641A..10P} took all the polarization and temperature information to test the possibility that the cosmic curvature power spectrum might be suppressed at superhorizon scales.  Even with polarization, cosmic variance limits what can be said: the data do not call for a suppression of power.

Our purpose here is to highlight and discuss the implications of a related curious feature in the large-angle CMB polarization, as seen in the leftmost side of Figs.~1 and 2 in Ref. \cite{Planck:2018vyg}, which show, respectively, the CMB temperature and polarization power spectra (and also the temperature-polarization cross-correlation).  The first Figure shows the longstanding low CMB temperature quadrupole, the feature that has led over the past thirty years to the speculation that something interesting may be happening at superhorizon scales.  The second Figure, however, shows something surprising:  The {\it polarization} quadrupole is {\it high}.  The occurrence of a low temperature quadrupole and a high polarization quadrupole warrants attention.  It is noteworthy because none of the other lowest multipole moments show such a discrepancy and even moreso if viewed as our best portal to superhorizon physics.

We quantify the discrepancy by calculating the joint probability distribution function (PDF) for the two quadrupoles and estimate that they are inconsistent at a $2.3\sigma$ level, a conclusion robust to changes in the cosmological model.  Given that the temperature quadrupole has now been obtained independently by three different satellite missions, we surmise that there may be issues with instrumental systematics and/or foreground removal in Planck.  If not, though, it suggests that new-physics explanations for the low temperature quadrupole are disfavored.  The discrepancy highlights the importance of LiteBIRD's improved measurements of these quadrupoles \cite{LiteBIRD:2022cnt}.

This paper is organized as follows: in Sections \ref{sec:multipoles} and \ref{sec:origins} we review the statistics of CMB multipole coefficients and their power spectra under the assumptions of $\Lambda$CDM cosmology, or more generally, a cosmological model with perturbations that are statistically isotropic/homogeneous and Gaussian.\footnote{A recent paper exploring the joint probability of four different CMB anomalies occurring together has called into question the assumption of statistical isotropy \cite{2023arXiv231012859J}. Our paper does not address this consideration.} In Section \ref{sec:discrepancy} we quantify the discrepancy between the temperature and polarization quadrupoles and their $\Lambda$CDM predictions; we show that the magnitude of this discrepancy grows when considering their joint probability. Superhorizon modifications to $\Lambda$CDM which would shift the quadrupole values are explored in Section \ref{sec:superhorizon}. We summarize our findings with a view toward LiteBIRD in Section \ref{sec:conclusion}.  

\section{CMB multipoles and power spectra}\label{sec:multipoles}

The standard $\Lambda$CDM cosmological model tells us that inflation produced adiabatic, nearly scale-invariant, and Gaussian primordial scalar (i.e. curvature) perturbations; these perturbations are the seeds that later induce CMB temperature and E-mode polarization anisotropies. The observed anisotropy patterns may be expanded in terms of spherical harmonics. For temperature anisotropies this takes the form,
\begin{equation}
    T(\mathbf{n}) = \sum_{l=1}^{\infty} \sum_{m=-l}^{l} a^T_{l m} Y_{l m}(\mathbf{n}), 
\end{equation}
which can be inverted to give harmonic coefficients,
\begin{equation}
    a^T_{l m} = \int d^2\mathbf{n} \, T(\mathbf{n}) Y_{l m}^*(\mathbf{n}).
\end{equation}
In the case of polarization anisotropies, we first decompose the sky pattern using Stokes $Q$ and $U$ parameters to get
\begin{equation}
    (Q \pm iU)(\mathbf{n}) = \sum_{l=1}^{\infty} \sum_{m=-l}^{l} a^{(\pm 2)}_{l m} \vphantom{|}_{(\pm 2)}Y_{l m}(\mathbf{n}),
\end{equation}
where $\vphantom{|}_{(\pm 2)}Y_{l m}(\mathbf{n})$ are the spin-2 spherical harmonic functions which are required since polarization is a spin-2 quantity. The associated harmonic coefficients are
\begin{equation}
    a^{(\pm 2)}_{l m} = \int d^2\mathbf{n} \, (Q \pm iU)(\mathbf{n})\vphantom{|}_{(\pm 2)}Y^*_{l m}(\mathbf{n}),
\end{equation}
which can easily be converted to E-mode coefficients
\begin{equation}
    a^{E}_{l m} = -\frac{1}{2} \left( a^{(+2)}_{l m} + a^{(-2)}_{l m} \right).
\end{equation}
Each $a_{l m}$ is drawn from a Gaussian with zero mean but nonzero variance---or angular power spectrum---$C_l$ given by,
\begin{equation}
    \left< a_{l m}^X a_{l' m'}^{Y*} \right> = \delta_{l l'} \delta_{m m'}C_l^{XY}, 
\end{equation}
where $X,Y \in \{T,E\}$.

\section{The origins of the quadrupoles}\label{sec:origins}

\begin{figure}
	\includegraphics[width=\columnwidth]{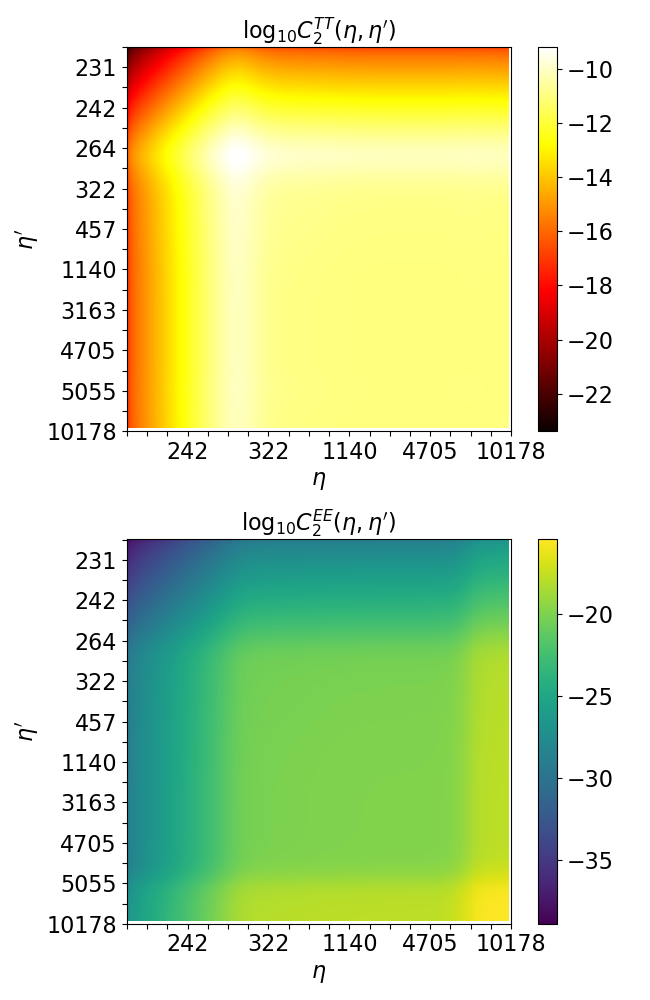}
    \caption{Logarithm of the conformal-time integrands of the temperature (top) and E-mode polarization (bottom) quadrupoles.}
    \label{fig:tt_ee_colormap}
\end{figure}

While both types of CMB anisotropies trace their origins to primordial curvature perturbations, the timescales over which their respective power spectra receive their greatest contributions differ. From a distribution of primordial curvature perturbations $\zeta(\mathbf{k})$ with power spectrum defined via
\begin{equation}
    \left< \zeta(\mathbf{k}) \zeta^*(\mathbf{k'}) \right> = P_\zeta(k) \delta(\mathbf{k} - \mathbf{k'}), 
\end{equation}
and dimensionless power spectrum $\mathcal{P}_\zeta(k) = k^3P_\zeta(k)/2\pi^2$, one can calculate the CMB angular power spectra,
\begin{equation}\label{eq:C_ell}
     C_l^{XY} = 4\pi \int \frac{dk}{k} \,\mathcal{P}_\zeta(k) \Delta^X_l(k,\eta_0) \Delta^Y_l(k,\eta_0),
\end{equation}
where $\Delta^X_l(k,\eta_0)$ are the photon transfer functions evaluated today. These can be obtained numerically, e.g., from a Boltzmann code like {\tt CLASS} \cite{Lesgourgues:2011re} or {\tt CAMB} \cite{Lewis:1999bs}. For the quadrupole, the E-mode transfer function can be approximated by \cite{Ji:2021djj}
\begin{equation}
     \Delta^E_2(k,\eta_0) \simeq -\frac{2\sqrt{6}}{3} \int_0^{\eta_0} d\eta\, g(\eta) \frac{j_2[k(\eta_0-\eta)]}{[k(\eta_0-\eta)]^2} j_2[k(\eta-\eta_*)],
\end{equation}
where $\eta_0$ is the conformal time today and $\eta_*$ that at recombination, and $g(\eta)$ is the visibility function for reionization; it integrates to the reionization optical depth $\tau= \int \, d\eta \, g(\eta)$ and peaks at the conformal time of reionization.  The temperature transfer function can be approximated by
\begin{equation}
     \Delta^T_2(k,\eta_0) \simeq -\frac{2}{9} j_2[k(\eta_0-\eta_*)].
\end{equation}
It is determined primarily by the quadrupole at the surface of last scatter and has an $O(\tau)$ correction (not shown) from reionization and also a small contribution from the integrated Sachs-Wolfe effect.   Fig.~\ref{fig:tt_ee_colormap} shows contributions to the CMB TT (top) and EE (bottom) power spectra as functions of $\eta$ and $\eta'$, using exact results from {\tt CLASS} \citep{Lesgourgues:2011re}. 

\begin{figure}
	\includegraphics[width=\columnwidth]{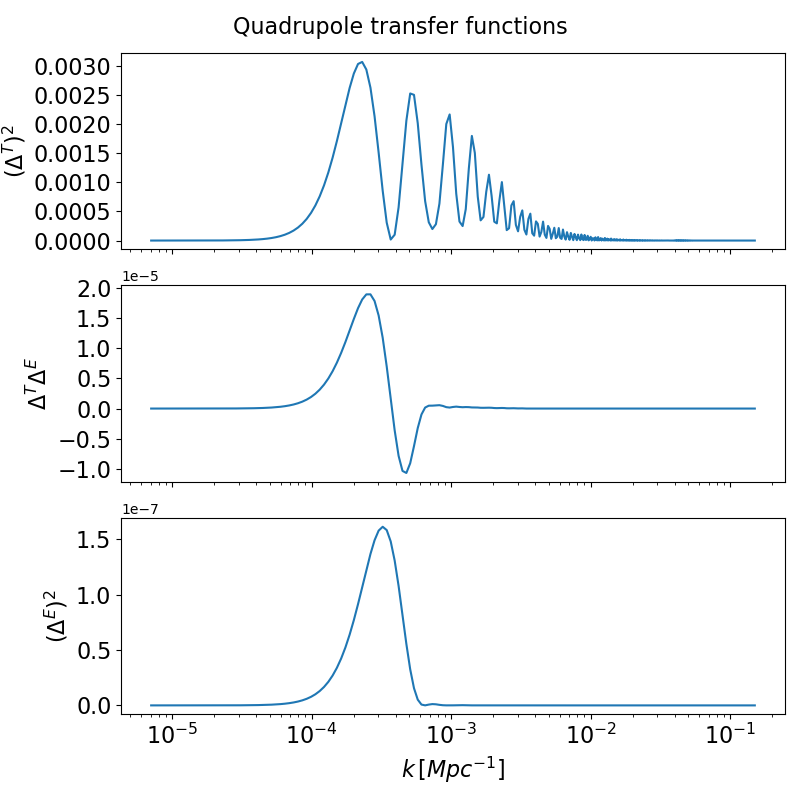}
    \caption{The CMB quadrupole transfer functions for the temperature (top), temperature-polarization (middle), and polarization (bottom) power spectra.}
    \label{fig:transfer_functions}
\end{figure}

Even though the peaks of recombination and reionization transfer functions occur at substantially different redshifts ($z_*\sim 1100$ and $z_{re} \sim 6$, respectively), the comoving distances to their surfaces are comparable. Consequently, the two probe a comparable range of physical separations, as shown in Fig.~\ref{fig:transfer_functions}.

\section{Discrepancy between quadrupoles and $\Lambda$CDM}
\label{sec:discrepancy}


\subsection{Noiseless, full-sky observations}

We can estimate the level of discrepancy of the measured moments
with the $\Lambda$CDM expectation analytically.
Our aim here is to assess {\it how much more} unusual the
low temperature quadrupole is, in light of the high polarization
quadrupole.  To do so, we make the simplifying assumption of noiseless measurements taken over a full-sky map.  The complications induced by noise and Planck's $86\%$ ($50\%$) sky coverage in temperature (polarization) \cite{2020A&A...641A...5P} will reduce the statistical significances but leave the basic conclusions unaltered. We will assess more precisely the impact of imperfect measurements and $f_{\rm{sky}} < 1$ in the next subsection.

If primordial perturbations are Gaussian, then each
spherical-harmonic coefficient $a_{lm}$ derived from a full-sky
map is a Gaussian random variable with zero mean and variance
$C_l$.  The estimators $\widehat{C_l^{TT}}$ and
$\widehat{C_l^{EE}}$ for the temperature/polarization moments
derived from a noiseless full-map are thus $\chi^2$-distributed variables
with $2l+1$ degrees of freedom.  Given the cross-correlation
$C_l^{TE}$ between the temperature and polarization
spherical-harmonic coefficients, though, the estimators
$\widehat{C_l^{TT}}$ and $\widehat{C_l^{EE}}$ are correlated for
any given $l$.  The joint PDF for $\chi^2$ variables 
correlated in this way can be related to the Wishart distribution \cite{2006MNRAS.372.1104P,2008PhRvD..77j3013H} and is derived in Appendix \ref{app:derivation}.  In our
case, the variables can be regarded as
$X=5 C_2^{TT,{\rm observed}}/C_2^{TT,{\rm expected}}$ and
$Y=5 C_2^{EE,{\rm observed}}/C_2^{EE,{\rm expected}}$, with $n=5$.  The
cross-correlation coefficient is $\rho =
C_2^{TE}/\left[C_2^{TT} C_2^{EE}\right]^{1/2}$.  From Planck Legacy Archive data\footnote{http://pla.esac.esa.int/pla/\#home} used by Figs.~1
and 2 in Ref. \cite{Planck:2018vyg}, we estimate this to be
$\rho\simeq 3/\sqrt{(1000)\times (0.055)}\simeq 0.5$.  Again, from these data, we infer a measured value of $Y\simeq 18$
(the polarization quadrupole is roughly three times its expected
value, $2l+1=5$) and $X\simeq
1.1$, 0.22 times the expected value

First, let's re-visit the temperature quadrupole.  The
cumulative distribution for a $\chi^2$-distributed variable $X$
with 5 degrees of freedom below one fifth the mean value is
about $4.5\%$, implying that a departure from the mean of this
magnitude will occur (taking into account the possibility of an
unusually high fluctuation) in roughly 1 out of every 11
trials, about a $1.7\sigma$ fluctuation.  
The fluctuation in the polarization quadrupole
is more significant---roughly a $2.7\sigma$ departure, a 1-in-150 occurrence.

If, however, we take into account the joint PDF for the two quadrupoles, the observed values of the two quadrupoles together
occur in roughly one in 25,000 realizations, a $4.1\sigma$ fluctuation.

\subsection{Imperfect measurements on a cut sky}

What we obtain from Planck is not the idealized picture of the microwave sky heretofore discussed. The measured power spectra contain contributions from residual foregrounds and instrumental systematics, and they are made on a masked sky map. Any statements regarding the statistical significance of CMB anomalies must account for these limitations.  Here we introduce a simple model to approximate the effects of partial-sky coverage and instrumental noise.
 
 To begin, we recall that for a full-sky map and no noise, the PDF for the observed temperature/polarization quadrupoles are given by $P(X,Y)$ with $X=5\, C_2^{TT,{\rm obs}}/C_2^{TT,{\rm exp}}$ and $Y=5\, C_2^{EE,{\rm obs}}/C_2^{EE,{\rm exp}}$, with $n=5$ and $\rho=C_2^{TE}/\left[C_2^{TT} C_2^{EE} \right]^{1/2}$.

Instrumental noise provides additional contributions $C_l^{TT,n}$ and $C_l^{EE,n}$ to the temperature and polarization power spectra, respectively, but none to the cross-correlation.  In practice, the temperature noise in Planck is negligible for the quadrupole and so we neglect it.  With these assumptions (still assuming full-sky coverage), the $a_{2m}^T$ are still distributed with variance $C_l^{TT}$.  The measured $a_{2m}^E$ include both signal and noise and are distributed with variance $C_2^{EE}+C_2^{EE,n}$.  The joint PDF for the observed quadrupoles is then $P(X,Y)$ where now $Y = 5 (C_2^{EE,{\rm obs}}+C_2^{EE,n})/(C_2^{EE}+C_2^{EE,n})$.  Note that here, $C_2^{EE,{\rm obs}} = C_2^{EE,{\rm meas}}-C_2^{EE,n}$; i.e., it is the true quadrupole we infer from the measured quadrupole after subtracting the expected noise contribution.  The cross-correlation coefficient is now reduced to $\rho=C_2^{TE}/\left[C_2^{TT} (C_2^{EE} +C_2^{EE,n})\right]^{1/2}$.

Now consider the effects of fractional sky coverage.  This would be easy if both the temperature and polarization maps had the same $f_{\rm sky}$.  In this case, we would replace $n=5\to 5\, f_{\rm sky}$ and also replace $5\to 5\, f_{\rm sky}$ in the definitions of $X$ and $Y$.  However, given that $f_{\rm sky}$ is different for polarization and temperature, a bit more thought is required.  To treat this, we first consider the temperature quadrupole.  The PDF for this can simply be taken to be a $\chi^2_{f_{\rm sky}^{T} (2l+1)}$ distribution.  We can then infer from Eq.~(\ref{eq:joint_pdf}) the conditional PDF, for our given $C_2^{TT}$, for $C_2^{EE}$ with $n$ scaled by $f_{\rm sky}^{E}$.

\begin{figure}
	\includegraphics[width=\columnwidth]{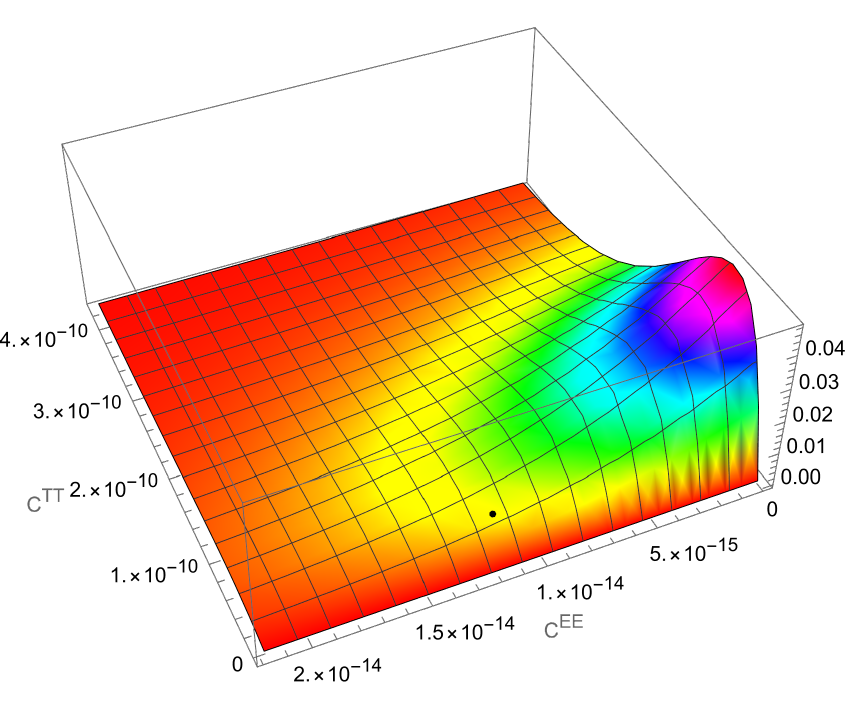}
    \caption{The unnormalized PDF of Eq.~(\ref{eq:joint_pdf_fsky}), ranging from lower values in red to higher values in purple. The black point corresponds to the observed temperature and polarization quadrupole values.}
    \label{fig:joint_pdf}
\end{figure}

To be more precise, we first note that
\begin{equation}
     P_n(X) \equiv \int_0^\infty\, dY\, P_n(X,Y;\rho),
\end{equation}
is a $\chi_n^2$ distribution for $X$.  We then note that the conditional probability for $Y$, given some value $X$, is
\begin{equation}
    P_n(Y;X,\rho) = \frac{P_n(X,Y;\rho)}{P_n(X)}.
\end{equation}
The unnormalized joint PDF for the temperature quadrupole $C_2^{TT,{\rm obs}}$ observed over $f_{\rm sky}^T$ of the sky and polarization quadrupole $C_2^{EE,{\rm obs}}$ observed over $f_{\rm sky}^E$ of the sky shown in Fig.~\ref{fig:joint_pdf} is

\begin{widetext}
\begin{equation}\label{eq:joint_pdf_fsky}
     \frac{P_{5\, f_{\rm sky}^T } \left( 5 \, f_{\rm sky}^T C_2^{TT,{\rm obs}}/C_2^{TT, {\rm exp}} \right) }{P_{5\, f_{\rm sky}^E } \left( 5 \, f_{\rm sky}^E C_2^{TT, {\rm obs}}/C_2^{TT, {\rm exp}} \right) } P_{5\, f_{\rm sky}^E} \left(  5 \, f_{\rm sky}^E \frac{ C_2^{TT,\rm obs}}{C_2^{TT,\rm exp}}, 5\, f_{\rm sky}^E \frac{ C_2^{EE,\rm obs} + C_2^{EE,n}}{ C_2^{EE,\rm exp} + C_2^{EE,n}}; \rho \right).
\end{equation}
\end{widetext}

The two-tailed probability of getting a temperature quadrupole at least as extreme as we observe, given Planck's limited sky coverage, is 12.8\% or a $1.5\sigma$ fluctuation. Table 3 of Ref. \cite{2020A&A...641A...5P} gives the probability of obtaining a polarization quadrupole as anomalous as measured as 29.6\%, or $1\sigma$, based on FFP10 simulations \cite{2020A&A...641A...3P}. From this result, we can solve for $Y$ using the equation
\begin{equation}\label{eq:get_noise}
    \int_Y^\infty \chi^2(5f_{\rm sky}^E, Y) \,dY = 0.296/2
\end{equation}
which gives $Y=4.6$. The amount of noise is thus $C_2^{EE,n}=4.1 \times 10^{-15}$. We now have all the ingredients required to calculate the joint probability. Integrating Eq.~(\ref{eq:joint_pdf_fsky}) and dividing by its normalization factor gives a two-tailed probability of 2.1\% or $2.3\sigma$.  

\section{A large-scale modification to $\Lambda$CDM}
\label{sec:superhorizon}

Taken at face value, the results of Section \ref{sec:discrepancy} are in minor conflict with the standard-cosmological-model expectations. We thus consider a modification to $\Lambda$CDM in which we modify the matter power spectrum $P_\mathrm{m}(k)$ at $k \lesssim 10^{-3}$ Mpc$^{-1}$. This appears in the low-$l$ CMB power spectra but not in large-scale structure surveys which probe higher $k$. For simplicity, we do not consider any effects that modifying the power spectrum will have on the inferred cosmological parameters.  

The full CMB quadrupole transfer functions are shown in Fig.~\ref{fig:transfer_functions}. From here we can see that even though temperature and E-mode polarization probe different epochs, the two extend over comparable distance scales. This provides us a $k$ range over which to invoke new physics and subsequently alter the matter power spectrum. 

We first consider decreasing $P_\mathrm{m}(k)$ by $P_\mathrm{m} \rightarrow P_\mathrm{m}/17$ for $k<3\times10^{-4}$ Mpc$^{-1}$. This is one way ensure that our modified $\Lambda$CDM model correctly predicts $C_2^{TT,\rm obs}$. The complete set of model predictions are $C_2^{TT} = 3.18\times 10^{-11}$, $C_2^{TE} = 5.52\times 10^{-14}$, and $C_2^{EE} = 3.17\times 10^{-16}$; this implies that the probability of our model generating an E-mode quadrupole at least as discrepant as $C_2^{EE,\rm obs}$ is 4.1\%. Put another way, observations and theory would be in tension at $2\sigma$. If instead we appropriately boost the large-scale power of our model to agree with $C_2^{EE,\rm obs}$, the probability of having an anomalous temperature quadruople comparable to or greater than what we see is 2.6\% or a $2.2\sigma$ discrepancy. We conclude that the tension between the two quadrupoles is not easily resolved by an increase/decrease of large-scale power.

\begin{figure}
	\includegraphics[width=\columnwidth]{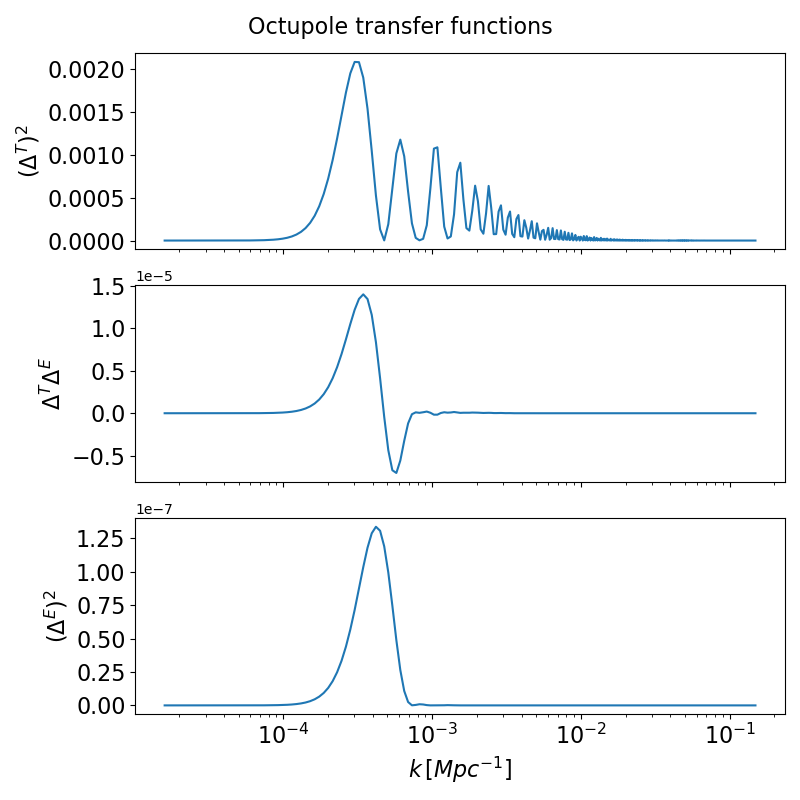}
    \caption{Same as Fig~\ref{fig:transfer_functions} but for the octupole moment.}
    \label{fig:octupole}
\end{figure}

\section{Conclusions}\label{sec:conclusion}

The low CMB temperature quadrupole has been a curiosity for three decades \cite{Bennett:1996ce}.  Although the inconsistency with its expected value is not that statistically significant, the fact that the power on the {\it largest} observable scale is low has been interpreted as a possible hint of new superhorizon physics.  The independent confirmation of a low quadrupole by WMAP \cite{WMAP:2012fli} and then Planck \cite{Planck:2018vyg} suggests that this low quadrupole is robust.

More recently, however, Planck has found the CMB polarization quadrupole to be somewhat high. Here we have calculated the joint PDF for the two quadrupoles and found that they are inconsistent with the $\Lambda$CDM expectation at the $2.3\sigma$ level.  We also argued that the inconsistency is robust to simple changes to the cosmological model.

What to make of this result?  It is obvious to wonder whether the Planck polarization result may be skewed high by some instrumental systematic and/or unusual foreground contamination. The former was mitigated a few years ago with the development of the {\tt SRoll2} algorithm \cite{2019A&A...629A..38D}, while the latter was considered more recently with the updated {\tt SRoll3} neural network \cite{2021A&A...651A..65L}. Results using the {\tt SRoll2} algorithm included a lower measured value for $C_2^{EE}$ with a PTE of 38\%, while the PTE for $C_2^{TE}$ decreased to 58.1\% \cite{2020A&A...635A..99P}. The temperature-polarization joint probability was not calculated, but we expect it to be similar to the PR3 results calculated in this work. 

Fortunately, though, COBE, WMAP, and Planck will not be our only sources of quadrupole-scale CMB measurements for long.  The forthcoming LiteBIRD experiment will target precise polarization measurements on the largest angular scales, with unprecedented frequency coverage to enable foreground removal over the entire sky. The anomaly should also motivate the search for some type of calibration of the polarization quadrupole, analogous to that suggested for the temperature quadrupole in Ref.~\cite{2003PhRvD..67f3001K} (the peculiar-velocity-induced temperature quadrupole discussed in there does not produce a polarization quadrupole).   If the marginally high polarization quadrupole persists, it will then disfavor explanations for the low quadrupole that involved suppressed large-scale power.  But will it then simply be chalked up to a statistical fluctuation?  Or perhaps some other new physics?

\section*{Acknowledgements}

We thank J.\ Dunkley for useful discussions.  This work was
supported at Johns Hopkins by NSF Grant No.\ 2112699 and the
Simons Foundation. JG acknowledges support from Princeton's Presidential Postdoctoral Research Fellowship.

\appendix

\section{Derivation of the joint $\chi^2$ probability
distribution function}\label{app:derivation}

Let $x$ and $y$ be Gaussian random variables with unit variance,
zero mean, and covariance $\vev{xy}=\rho$.  If we have $n$
realizations $x_i$ and $y_i$ (for $i=1,\ldots,n$) of these
pairs, then the joint probability distribution function (PDF)
for these $2n$ random variables is
\begin{equation}
     p(\vec x,\vec y) = \frac{1}{(2\pi)^n (1-\rho^2)^{n/2}} \exp
     \left[ - \frac{\vec x^2 +\vec y^2 - 2 \rho \vec x
     \cdot \vec y}{2 (1-\rho^2)} \right],
\end{equation}
where $\vec x$ and $\vec y$ are $n$-dimensional vectors with
components $x_i$ and $y_i$, respectively.

We now define two variables $X \equiv \vec x^2$
and $Y \equiv \vec y^2$ which are correlated $\chi^2$ variables
with joint PDF,
\begin{equation}
     P_n(X,Y;\rho) = \int d^nx \int d^ny\, p(\vec x,\vec y) \delta_D(X-\vec x^2)
     \delta_D(Y-\vec y^2),
\label{eqn:PXY}     
\end{equation}
where $\delta_D(x)$ is the Dirac delta function.  The
probability distribution is a function of the vector norms $\vec
x^2$ and $\vec y^2$, and also the dot product $\vec x \cdot \vec 
y$, and we are integrating over all $\vec x$ and all $\vec y$.
The integration over $\vec x$ can be written as $S_{n-1} \int
x^{n-1}\,dx$, where $S_{n-1} = 2 \pi^{n/2}/\Gamma(n/2)$ is the
volume of the unit $n$-sphere.  The integration over $\vec y$
can then be written as $S_{n-2} \int \, dy_1 \int y_\perp^{n-2}\,
dy_\perp$, where $y_1$ is the component of $\vec y$ along the
direction of $\vec x$ and $\vec y_\perp=(y_2,\ldots, y_n)$ the
components perpendicular to $\vec x$.  We then write $x^{n-1}\, dx = (x^2)^{(n/2)-1}\, d(x^2)/2$, $y_\perp^{n-2}\, dy_\perp = (y_\perp^2)^{(n/2)-(3/2)}\, d(y_\perp^2)/2$, and the argument of the second Dirac delta function in Eq.~(\ref{eqn:PXY}) as $(Y-y_1^2)-y_\perp^2$.  
We then find
\begin{equation}\label{eq:joint_pdf}
\begin{aligned}
     P_n(X,Y;\rho) &= \frac{(XY)^{\frac{n}{2}-1}}{2^n (1-\rho^2)^{n/2}
     \left[\Gamma(n/2) \right]^2} \exp\left[ -\frac{X+Y}{2
     (1-\rho^2)} \right] \\ &\times G_n\left( \frac{\rho
     \sqrt{XY}}{1-\rho^2} \right),
\end{aligned}
\end{equation}
where
\begin{equation}
     G_n(\alpha) \equiv \frac{1}{\sqrt{\pi}} \frac{
     \Gamma\left(\frac{n}{2} \right)}{\Gamma\left( \frac{n-1}{2}
     \right)} \int_{-1}^1\, d\mu \, (1-\mu^2)^{\frac{n}{2}
     -\frac32} e^{\alpha \mu}.
\end{equation}     
 Mathematica can write the integral in terms of a hypergeometric function, which provides little insight but may facilitate numerical evaluation.

\bibliography{apssamp}

\end{document}